\documentclass[reprint,amsmath,amssymb,aps,prb,showpacs]{revtex4-1}
\usepackage[T1]{fontenc}

\usepackage{graphicx}
\graphicspath{ {figures/} }
\usepackage{amsmath}
\usepackage{dsfont}

\usepackage[colorlinks,hyperindex]{hyperref}
\usepackage{color}

\begin{document}

\title{Spontaneous formation of spin lattices in semimagnetic exciton-polariton condensates}

\author{Pawe{\l} Mi\k{e}tki}
\author{Micha{\l} Matuszewski}
\affiliation{Institute of Physics, Polish Academy of Sciences, Al. Lotnikow 32/46, PL-02668 Warsaw, Poland}
  
\begin{abstract}
  An exciton-polariton microcavity that incorporates magnetic ions can exhibit a spontaneous self-trapping phenomenon which is an analog of the classical polaron effect. We investigate in detail the full model of a polariton condensate that includes pumping and losses, the spin degree of freedom, external magnetic field and energy relaxation. In the quasi-one-dimensional case, we show that the polaron effect can give rise to a spontaneous lattice of perfectly arranged polarization domains in an antiferromagnetic configuration. We find that partial polarization of the condensate at moderate magnetic field strengths facilitates the formation of such ``polaron lattices'', which are qualitatively different from self-trapped polarons that appear in a fully polarized condensate. Within the Bogoliubov-de Gennes approximation, we calculate the instability condition which marks the appearance of the patterns. Surprisingly, we find that the stability condition displays a discontinuity at the point of partial-full polarization threshold.
\end{abstract}

\maketitle
%
%
%
%
\section{\label{sec:intro}Introduction}
%

Diluted magnetic (also known as semimagnetic) semiconductors are characterized by the exchange interaction between spins of magnetic ions and carriers, which leads to phenomena such as the giant Zeeman effect~\cite{Kossut_Review,Dietl_Review,Ivchenko,Kavokin_CoherentSpinDMS,Mirek_AngularDependence}. 
Magnetic polarons are spin-organized bound states formed due to this interaction.
This concept was first proposed by De Gennes in 1960~\cite{Gennes_Polaron} and thoroughly investigated both theoretically and experimentally in semimagnetic semiconductors~\cite{Dietl_Polaron,Mauger_Polaron,Goryca_SpinRelaxationQD,Dietl_DynamicsSpinOrganization,Dietl_SpinDynamicsConfinedElectron,ALaGuillame_FreeMagneticPolaron,Kavokin_Theory2DPolarons} in the cases of both impurity-bound and free (self-trapped) polarons.

In microcavity semiconductor structures, exciton-polariton quasiparticles exist when the exciton-photon coupling is strong enough~\cite{Hopfield_Polaritons,Weisbuch_Polaritons,Kavokin_Microcavities}.
These light-matter quasiparticles can Bose condense even at room temperature due their effective mass which is many orders of magnitude smaller than the electron mass~\cite{Kasprzak_BEC,Grandjean_RoomTempLasing,Kena_NonlinearOrganic}. 
Furthermore, exciton-polariton condensates allowed for the to observation of some fascinating phenomena from superfluid excitations~\cite{Amo_Superfluidity,Sanvitto_Superfluidity,Pietka_Josephson,Amo_Josephson,Deveaud_QuantumVortices,Sanvitto_PersistentCurrents,Deveaud_VortexDynamics}, to solitons~\cite{Amo_DarkSolitons,Sich_solitons,Sanvitto_nasz}. 
Several possible applications have been put forward as well, ranging from low threshold lasers~\cite{Yamamoto_NPRev, Kavokin_Laser}, to all-optical transistors~\cite{Sanvitto_Transistor,Savvidis_TransistorSwitch,Shelykh_Neurons}, to quantum simulation~\cite{Lagoudakis_XYModel,Yamamoto_QuantumSimulators}.

In semimagnetic polariton systems, it was demonstrated theoretically that self-trapping phenomenon can occur for realistic system parameters thanks to the strong exciton-ion interaction and the polariton coherence in the condensed state. Existence of self-trapped ``polariton-polarons'' was theoretically predicted both in the equilibrium case~\cite{Kavokin_InterplaySuperfluidityDMS} and in the non-equilibrium case which includes the effect of pumping and losses~\cite{Mietki_MagneticPolaron}. Note that qualitatively different, nonmagnetic collective polaron effect was observed in exciton-polarion system~\cite{Sanvitto_nasz} due to interaction with lattice phonons~\cite{Kavokin_HeatAssisted}.

In this paper, we investigate the magnetic self-trapping in a semimagnetic  polariton condensate taking into account both the spin degree of freedom, pumping and losses, and energy relaxation. We consider a Cd$_{1-x}$Mn$_x$Te microcavity that has been recently realized experimentally~\cite{Pacuski_StrongCoupling,Mirek_AngularDependence,Pacuski_MagneticFieldEffect}.
In our model, the magnetic ion subsystem is fully thermalized, but the polariton subsystem is far from thermal equilibrium, as suggested by experiments~\cite{Mirek_AngularDependence,Krol_SpinPolarized}. Nevertheless, we find that in the phase diagram of the system the inverse of polariton relaxation rate plays a role similar as an effective polariton temperature.

We show that the system spontaneously forms intricate spin structures even at relatively low magnetic field strength. We find that spontaneous spin lattices are formed with side-by-side antiferromagnetic arrangement of spin domains. At higher magentic fields or when the ion-exciton interaction is stronger, the system develops more typical polartion self-trapping similar as in the previously considered spin-polarized case~\cite{Kavokin_InterplaySuperfluidityDMS,Mietki_MagneticPolaron}. Using the Bogoliubov-de Gennes method, we calculate an analytical condition for stability of the system and compare it with numerical results.
Interestingly, we find a jump of the stability threshold when entering the spin-polarized state, which is due to the lack of partially spin polarized excitations in this case. Our results should pave the way for the first direct observation of magnetic self-trapping and pattern formation in a semiconductor system.

%
%
%
%

\section{\label{sec:model}Model}
%
We take into consideration a two-dimensional cavity with a microwire that confines the condensate in one dimension~\cite{Bloch_ExtendedCondensates,Deveaud_Disorder1D}.
In the mean field approximation, exciton-polaritons can be described with the coupled one-dimensional complex Ginzburg-Landau equations for the macroscopic wavefunctions~\cite{Mietki_MagneticPolaron,Kavokin_InterplaySuperfluidityDMS}
%
\begin{gather}
\begin{aligned}
i (1+ i\Gamma) \hbar \frac{\partial \psi_{\sigma}}{\partial t} &=  - \frac{\hbar^2}{2 m^*} \frac{\partial^2  \psi_\sigma}{\partial x^2} + g_1 |\psi_\sigma|^2 \psi_\sigma  + g_2 |\psi_{-\sigma}|^2 \psi_\sigma \\
+ iP\psi_\sigma - &i\frac{1}{2}\gamma_{\rm L} \psi_\sigma - i\gamma_{\rm NL} |\psi_\sigma|^2 \psi_\sigma - \sigma \lambda M \psi_\sigma \label{eq:CGLE1} 
\end{aligned}
\end{gather}
%
where $(1 + i\Gamma)$ is a term that corresponds to energy relaxation with the energy dissipation factor $\Gamma$~\cite{Pitaevskii_EnergyRelaxation}. This term introduces not only relaxation of kinetic energy, but also relaxation in the spin space between two polarizations $\sigma=\sigma_+$, $\sigma_-$. The $g_1$ and $g_2$ coefficients are constants of interaction between same- and oppositely-polarized spins,
$P$ is the external uniform pumping, $m^*$ is the effective mass, and $\gamma_{\rm L}, \gamma_{\rm NL}$ are linear and nonlinear loss coefficients.
The last term corresponds to the influence of diluted magnetic ions. This effective additional potential depends on spin $\sigma$, magnetic ion-polariton interaction constant $\lambda$ and the mean-field ion magnetization $M(x,t)$. Note that in our simple model, we do not take into account the exciton reservoir as a separate degree of freedom. Such assumption is justified in the limit of adiabatic approximation to the reservoir dynamics~\cite{Bobrovska_Adiabatic,Bobrovska_Stability}.

Magnetic ion dynamics can be described by the spin relaxation equation~\cite{Kavokin_Polaron}
%
\begin{gather}
\frac{\partial M(x,t)}{\partial t} = \frac{ \langle M(x, t) \rangle - M(x,t)}{\tau_{\rm M}} \label{eq:MTR1}
\end{gather}
%
with a characteristic ion spin relaxation time $\tau_{\rm M}$. Here, $\langle M(x, t) \rangle$ is the equilibrium value of magnetization given by the Brillouin function~\cite{Gaj_Brilloiun}
%
\begin{gather}
	\langle M(x, t) \rangle= n_{\rm M} g_{\rm M} \mu_{\rm B} J~B_{\rm J}\left( \frac{g_{\rm M} \mu_{\rm B} J B_{\rm eff}}{k_{\rm B} T} \right)  \label{eq:Mmean}
\end{gather}
%
where $n_{\rm M}$ is the concentration of ions, $g_{\rm M}$ is the \textit{g}-factor, $J$ is the total spin of a Mn ion, $\mu_{\rm B}$ is the Bohr magneton, $T$ is the temperature of the ion subsystem. 
The magnetic field felt by the ions is effectively increased by the spin polarization of the condensate
%
\begin{gather}
	B_{\rm eff} = B_{0} + \lambda S_{\rm Z} \label{Beff1},
\end{gather}
%
where  $B_0$ is the external magnetic field. The pseudospin density $S_{\rm Z}$ is given by $\frac{1}{2}(|\psi_+|^2 - |\psi_-|^2)$ and polariton-ion coupling constant $\lambda$ is given by the ion-exciton exchange interaction $\beta_{\rm EX}$, the excitonic Hopfield coefficient $X$ and the width of the quantum well $L_{\rm Z}$~\cite{Kavokin_InterplaySuperfluidityDMS}
%
\begin{gather}
	\lambda = \frac{\beta_{\rm EX} X^2}{\mu_{\rm B} g_{\rm M} L_{\rm Z}}.
\end{gather}
%
We neglect the effect of intrinsic exciton Zeeman splitting that is unnoticeable at weak fields~\cite{Pietka_MagneticFieldTuning,Mirek_AngularDependence} and TE-TM splitting which could cause polariton spin precession~\cite{Shelykh_PolarizationPropagation}, but can be avoided by an appropriate sample design.

%
%
%
%
\section{\label{sec:stat_sol}Homogeneous solutions}
%

%
\begin{figure}
	\includegraphics[width=0.5\textwidth]{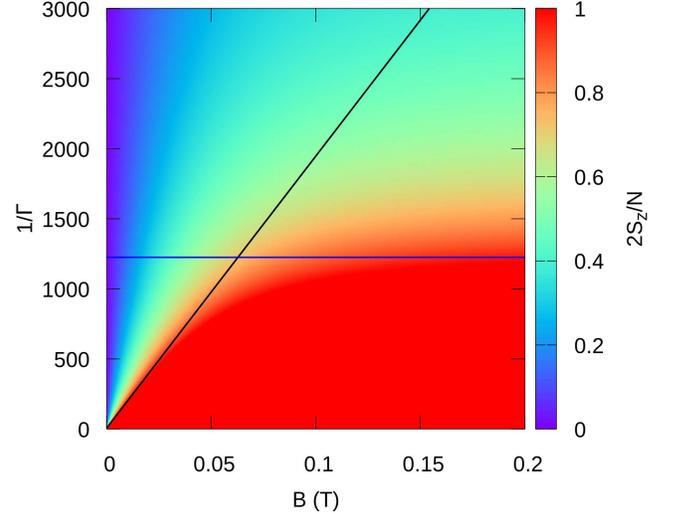}
    \caption{Pseudospin polarization degree of a homogeneous condensate shown in coordinates of the magnetic field $B$ and the inverse of energy relaxation $1/\Gamma$. Black and blue lines are theoretical boundaries of full polarization ($S_z=N/2$) in the weak and strong magnetic field limit, respectively. The dependence of polarization degree on $1/\Gamma$ allows to interpret it as an effective nonequlibrium ``temperature''.}
    \label{fig:gamma-B}
\end{figure}
%
%
We begin the analysis of the system by considering stationary homogeneous states in the absence of self-trapping.
The stationary solutions can be described with the density $n_{\rm \sigma}$ and the chemical potential $\mu_{\rm \sigma}$ of each component
%
\begin{gather}
	\psi_+^{(0)}(x,t) = \sqrt{n_{\rm +}} e^{-i\mu_+ t/\hbar} \label{eq:stat+} \\ 
    \psi_-^{(0)}(x,t) = \sqrt{n_{\rm -}} e^{-i\mu_- t/\hbar} \label{eq:stat-} \\
    M^{(0)}(x,t) = \langle M \rangle 
    \label{eq:statM}
\end{gather}
%
%
After substituting  Eqs.~(\ref{eq:stat+}) and~(\ref{eq:stat-}) into~(\ref{eq:CGLE1}), from the real and imaginary part of the equation we obtain the conditions
%
\begin{align}
P_{\rm eff} - n_{\rm +} \gamma_{\rm NL} - \Gamma (n_{\rm +} g_1 + n_{\rm -} g_2 - \zeta \lambda M)  = 0 \label{eq:Im1},\\
P_{\rm eff} - n_{\rm -} \gamma_{\rm NL} - \Gamma (n_{\rm -} g_1 + n_{\rm +} g_2 + \zeta \lambda M)   = 0 \label{eq:Im2},
\end{align}
%
where the effective pumping $P_{\rm eff} = P - \frac{1}{2}\gamma_{\rm L}$. Clearly, the terms in the bracket correspond to modifiction of losses due to relaxation, proportional to the potential for a given spin component.

Equations~(\ref{eq:Im1}), (\ref{eq:Im2}) together with Eq.~(\ref{eq:Mmean}) allow to find numerically densities $n_{\rm +}$, $n_{\rm -}$ and magnetizaton $M$.
In Fig.~\ref{fig:gamma-B} we show the polariton pseudospin polarization degree as a function of the magnetic field $B$ and the inverse of the energy relaxation $1/\Gamma$. 
The results were obtained by simulating system evolution without the kinetic energy term until a stable state was reached, for each point in the Figure. The computed mean value of the polarization degree i.e. $2 S_{\rm Z}/N$, where $N=|\psi_+|^2 + |\psi_-|^2$, is shown for the final steady states. 
We also depict in Fig.~\ref{fig:gamma-B} the analytically predicted boundaries of full polarization of steady states in the limit of weak nad strong magnetic field, marked with lines. 
In the weak magnetic field limit, the Brillouin function can be linearized and in the strong magnetic field limit, the value of magnetization saturates, so in both cases the equations become solvable. The details of the calculation are given in Appendix \ref{app:Wmf}.

We note that the phase diagram depicted in Fig.~\ref{fig:gamma-B} resembles the one that was obtained in the case of thermal equilibrium~\cite{Kavokin_InterplaySuperfluidityDMS} provided that the temperature is replaced with the inverse of the relaxation rate $1/\Gamma$.
Hence, one can argue that $1/\Gamma$ plays the role of an effective temperature of the polariton subsystem. Similar conclusions were obtained previously in several works discussing this analogy in the context of nonequilibrium condensates~\cite{Richard_ThermalDecoherence,Carusotto_Quasi-condensates,Carusotto_Laser,Carusotto_Pseudothermalization}. This analogy can be explained intuitively: for large relaxation rates $\Gamma$, as compared to the polariton lifetime, the system is expected to be close to the polariton ground state, which is the condensate state within the equilibrium theory. With decreasing temperature (or increasing relaxation rate) the ions and polaritons are more likely to align in the direction of the external magnetic field, which translates to a larger polarization degree.

%
%
%
%
\section{\label{sec:num_results}Instability and polaron formation}

%
\begin{figure}
	\includegraphics[width=0.5\textwidth]{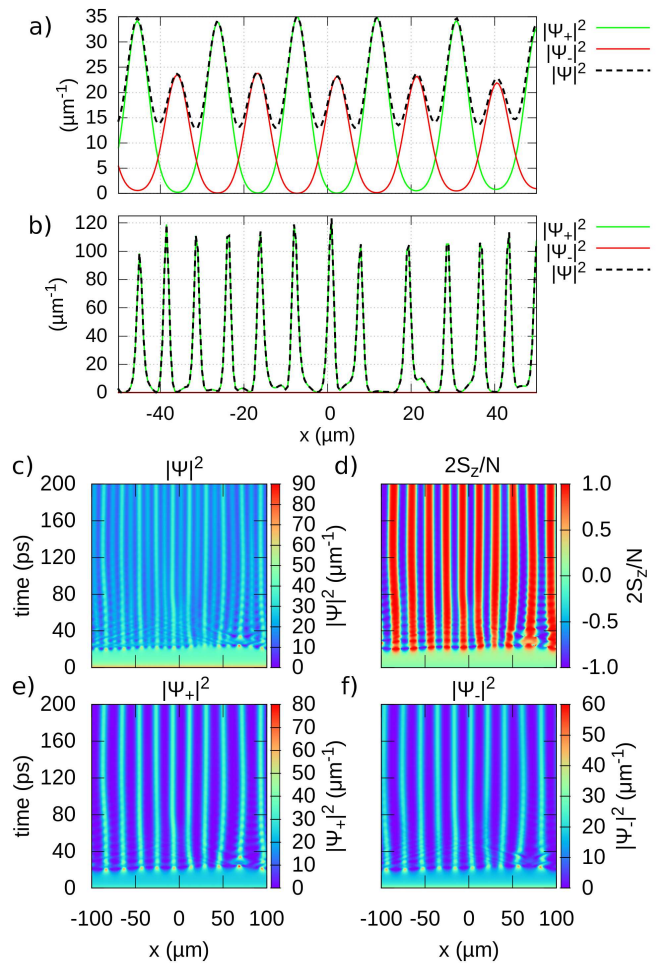}
        \caption{(a) Example of the densities of the $\sigma_+$ and $\sigma_-$ components and the total density $|\psi|^2=|\psi_+|^2+|\psi_-|^2$ in a ``polaron lattice'' state with alternating spin-up and spin-down domains. (b) The same for a fully polarized case, a set of localized polarons is visible.
          (c)-(f) Evolution of the exciton-polariton condensate leading to the formation of lattice from panel (a).
          Shown are (c) total density $|\psi|^2$,  (d) normalized pseudospin $S_{\rm Z}/|\psi|^2$, (e) density of the $\sigma_+$ component and (f) density of the $\sigma_-$ component. Parameters of the simulation are given in~\cite{Parameters}.}
    \label{fig:1d2s}
\end{figure}
%

%
We now investigate the stability of homogeneous states and demonstrate the formation of polarons and polaron lattices in the unstable regime. We take into account the spin degree of freedom, in contrast to previous studies where self-trapped polarons were fully polarized~\cite{Kavokin_InterplaySuperfluidityDMS,Mietki_MagneticPolaron}.
The self-trapping effect was shown to occur far from the thermal equilibrium~\cite{Mietki_MagneticPolaron} due to ion-exciton interaction, which induces an effective attractive interaction between polaritons. Within this interpretation, self-trapped polarons can be considered as bright solitons in analogy to the conservative nonlinear Schr\"odinger equation systems~\cite{Infeld_NonlinearWaves}.

Here, we show that in the case when the condensate is not fully polarized, the system can develop coherent spatial structures that are qualitatively different from such ``bright soliton'' polarons. They take the form of ``polaron lattices'', which are perfectly aligned domains of condensate polarization in an antiferromagnetic configuration, see Fig.~\ref{fig:1d2s}(a). Formation of these structures appears to be triggered by phase separation between spin-up and spin-down components, as follows from the analysis within the Bogoliubov approximation, described in detail in Sec.~\ref{sec:BdGapprox}. For comparison, in Fig.~\ref{fig:1d2s}(b) we show the ``bright soliton'' polaron structures that appear in the strong magnetic field regime, when the condensate is completely spin-polarized. Clearly, the arrangement of polarons in this case is less regular, and they differ in width and amplitude. The dynamics of such strudtures was described in our previous work~\cite{Mietki_MagneticPolaron}. 

In Figures~\ref{fig:1d2s}(c)-(f) we depict the typical dynamics of the system described by Eqs.~(\ref{eq:CGLE1}) and~(\ref{eq:MTR1}) in the case corresponding to Fig.~\ref{fig:1d2s}(a). The initial state is a stationary state as in Eqs.~(\ref{eq:stat+}) and~(\ref{eq:stat-}) disturbed by a small white noise.
The creation of ``polaron lattice'' appears to follow the same path as in the case of polarized polarons~\cite{Mietki_MagneticPolaron}, however with an important difference that the final state is of perfectly aligned and equal amplitude peaks.
The mean distance between  peaks is 
inversely proportional to the most unstable $k$-mode, i.e.~momentum that correspond to the maximum value of the imaginary branch of the Bogoliubov dispersion relation.
The total density, depicted in Fig.~\ref{fig:1d2s}(c) is only slightly varying. On the other hand, the polarization degree  in Fig.~\ref{fig:1d2s}(d) is strongly modulated due to the antiferromangetic configuration of domains. 
Importantly, such alternating spin structure can be the factor that will allow to distinguish self-localized polaron lattices from density fluctuations that are simply trapped in a defect of the sample.

%
\begin{figure}
	\includegraphics[width=0.5\textwidth]{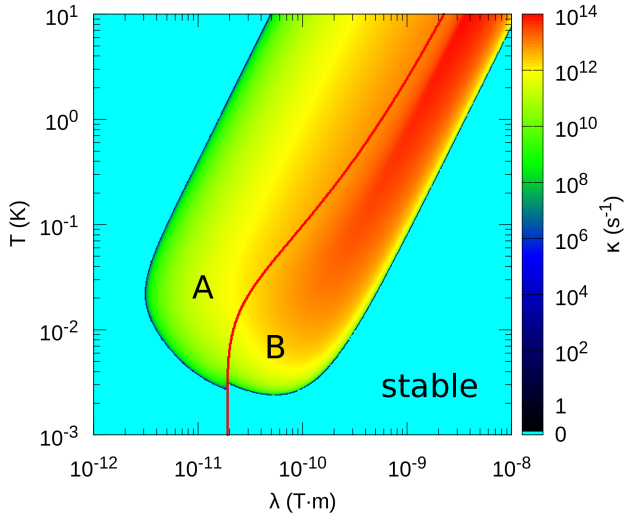}
    \caption{Diagram of stability in coordinates of the ion-polariton coupling constant $\lambda$ and temperature $T$. Red line corresponds to the boundary between fully and partially spin polarized condensate as predicted by the homogeneous state analysis. Polaritons are partially polarized on the left-hand side and fully on the right-hand side. The regions marked \textsf{A} and \textsf{B} corespond to the states shown in Fig.~\ref{fig:1d2s}(a) and~(b), respectively. Magnetic field $B$ = 0.01 T and relaxation $\Gamma$ = 0.001.}
    \label{fig:T-L}
\end{figure}
%

%
\begin{figure}
	\includegraphics[width=0.5\textwidth]{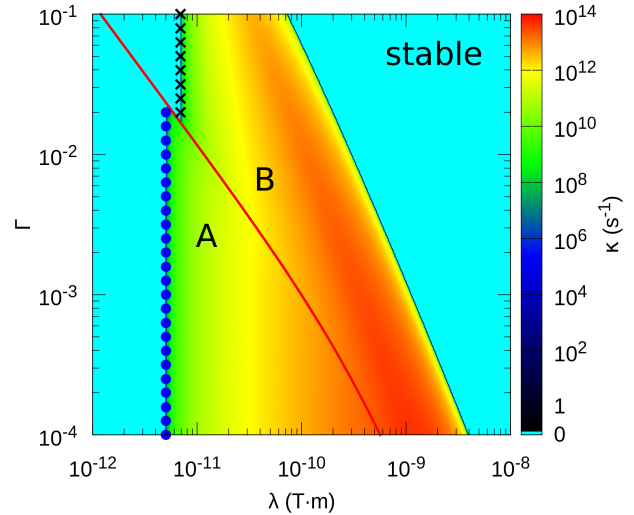}
    \caption{Same as Fig.~\ref{fig:T-L}, but in coordinates of the ion-polariton coupling constant $\lambda$ and energy relaxation factor $\Gamma$. Crosses and dots mark the theoretical predictions of stability treshold for fully and partially polarized condensate, Eqs.~(\ref{eq:condition1s}) and~(\ref{eq:condition2s}), respectively. Magnetic field $B$ = 0.01 T and temperature $T$ = 0.1 K}
    \label{fig:gamma-L}
\end{figure}
%

The crucial parameter for the emergence of polarons is the ion-polariton coupling that should be within an appropriate range. Weak coupling will not lead to a sufficiently strong ion mediated interaction effect, while too strong coupling leads to the saturation of the Brillouin function. 
Figures~\ref{fig:T-L} and~\ref{fig:gamma-L} present stability diagrams computed using the Bogoliubov-de Gennes method and verified numerically by solving Eqs.~(\ref{eq:CGLE1}) and~(\ref{eq:MTR1}). The figures are depicted
in parameter space of ion-polariton coupling $\lambda$ vs temperature $T$ and energy relaxation factor $\Gamma$, respectively. 
The color scale illustrates the instability rate: cyan color shows that the system is stable (it is marked as additional zero on the logarithmic scale).
Note that homogeneous states are partially polarized on the left side of the red line and fully polarized on the right side. Hence, the red line shows the boundary between partially and fully spin polarized condensate, although for the inhomogeneous polaron states the limit is slightly different than the analytical one depicted by the line.
One can observe that there is a non-continuous shift of stability threshold when crossing the red line. While this shift may seem tiny, one should take into account that the figures are plotted on a logarithmic scale. The shift of the stability threshold is actually quite substantial (about a factor of $\sqrt{2}$ on the $\lambda$ axis) and it is discussed in detail in Sec.~\ref{sec:BdGapprox}.

Note that in Fig.~\ref{fig:T-L} at very low temperatures the condensate is stable for all values of $\lambda$. 
As we previously demonstrated~\cite{Mietki_MagneticPolaron}, the range of such stable temperatures increases with the external magnetic field strength.
According to Fig.~\ref{fig:T-L} stability depends strongly on the temperature while in Fig.~\ref{fig:gamma-L}, for partially polarized condensate (left of the red line) stability does not depend on $\Gamma$ (see also Section~\ref{sec:BdGapprox}). Hence,  with regard to stability, the temperature of the ion sybsystem appears to be more important than the effective nonequilibrium temperature of the polariton subsystem. This is understandable as the response given by the Brillouin function depends explicitly on the ion temperature only. 
The crosses and dots in Fig.~\ref{fig:gamma-L} mark the analytical predictions of the stability boundary in the case of fully and partially polarized condensate, according to the Eqs.~(\ref{eq:condition2s}) and~(\ref{eq:condition1s}), which agree very well with the numerical results.

%
%
%
\section{\label{sec:BdGapprox}Stability analysis}
%
We perform analysis of stability of the condensate within the Bogoliubov-de Gennes approximation.
For convenience, we introduce a dimensionless form of the model. By rescaling space, time, wavefunction and other parameters as $x =\xi \tilde{x}$, $t = \alpha \tilde{t}$, $\psi_\sigma = (\xi\beta)^{-1/2} \tilde{\psi_\sigma}$, $g_{(1,2)} = \hbar\xi\beta\alpha^{-1} \tilde{g_{(1,2)}}$, $P_{\rm eff} = \hbar\alpha^{-1} \tilde{P}_{\rm eff}$, $\gamma_{\rm NL} = \hbar\xi\beta\alpha^{-1} \tilde{\gamma_{\rm NL}}$, $M = \zeta \tilde{M}$, $\lambda = \hbar\alpha^{-1} \tilde{\lambda}$, we obtain (we omit tildes below)
%
\begin{gather}
\begin{aligned}
i (1+ i\Gamma) &\frac{\partial \psi_{\sigma}}{\partial t} = - \frac{\partial^2  \psi_\sigma}{\partial x^2} + g_1 |\psi_\sigma|^2 \psi_\sigma  + g_2 |\psi_{-\sigma}|^2 \psi_\sigma \\
&+ iP_{\rm eff}\psi_\sigma - i\gamma_{\rm NL} |\psi_\sigma|^2 \psi_\sigma - \sigma \zeta \lambda M \psi_\sigma \label{eq:CGLE2}
\end{aligned}
\\ \frac{\partial M}{\partial t} = \frac{\alpha}{\tau_{\rm M}} \left[ J B_{\rm J}\left(\delta \lambda (|\psi_+|^2 - |\psi_-|^2)  \right) - M\right]
\end{gather}
%
where $\xi = \sqrt{\hbar\alpha/2m^*}$, $\zeta = g_{\rm M} \mu_{\rm B} n_{\rm M}$, $\delta = \frac{g_{\rm M} \mu_{\rm B} J}{2 k_{\rm B} T} \frac{\hbar}{\alpha \beta \xi }$, while $\alpha$, $\beta$ are free parameters of the scaling.

As we previously demonstrated~\cite{Mietki_MagneticPolaron}, the appearance of the polarons is related to the instability of the homogeneous stationary state.
To analyze the stability we perturb the stationary solution~\cite{Wouters_ExcitationSpectrum} (\ref{eq:stat+})-(\ref{eq:statM})
%
\begin{gather}
\begin{aligned}
	\psi_+ = \psi_+^{(0)} &\left[1 + \epsilon \sum_k \left\{ u_k(t) e^{ikx} + v_k(t) e^{-ikx} \right\} \right]  \\
    \psi_- = \psi_-^{(0)} &\left[1 + \epsilon \sum_k \left\{ r_k(t) e^{ikx} + s_k(t) e^{-ikx} \right\} \right]  \\
     M = M^{(0)} &\left[1 + \epsilon \sum_k \left\{ w_k(t) e^{ikx} + w_k^*(t) e^{-ikx} \right\} \right], \label{eq:perturbations1}
\end{aligned}
\end{gather}
%
where $\epsilon$ is a small parameter.
Substituting Eqs.~(\ref{eq:perturbations1}) into Eqs.~(\ref{eq:CGLE1}), (\ref{eq:MTR1}) and then taking $\epsilon$ up to the first order and expanding Brillouin function up to the first order term we obtain the usual eigenvalue problem $Q_k U_k = \omega_k U_k$ where $U_k = (u_k, v_k^*, r_k, s_k^*, w_k)^T$ and 
%
\begin{widetext}
\begin{align} \label{eq:Qmatrix}
    Q_k = 
\begin{pmatrix} 
	\left( k^2 + n_{\rm +} \tilde{g_1} \right)  \tilde{\Gamma} & 
      n_{\rm +} \tilde{g_1}             		\tilde{\Gamma} &
	  n_{\rm -} g_2 	 						\tilde{\Gamma} & 
	  n_{\rm -} g_2  							\tilde{\Gamma} & 
     -J B_{\rm J} \lambda \zeta  				\tilde{\Gamma} \\[0.3cm]
	 -n_{\rm +} \tilde{g_1}^*           		\tilde{\Gamma}^* &
    -\left(k^2 + n_{\rm +} \tilde{g_1}^* \right) \tilde{\Gamma}^* & 
	 -n_{\rm -} g_2 	 						\tilde{\Gamma}^* & 
	 -n_{\rm -} g_2 	 						\tilde{\Gamma}^* & 
      J B_{\rm J} \lambda \zeta  				\tilde{\Gamma}^* \\[0.3cm]
      n_{\rm +} g_2 	 						\tilde{\Gamma} & 
      n_{\rm +} g_2 	 						\tilde{\Gamma} & 
	\left( k^2 + n_{\rm -} \tilde{g_1}  \right) \tilde{\Gamma}	& 
      n_{\rm -} \tilde{g_1}           		    \tilde{\Gamma}	&
      J B_{\rm J} \lambda \zeta  				\tilde{\Gamma} \\[0.3cm]
     -n_{\rm +} g_2 	 						\tilde{\Gamma}^* & 
     -n_{\rm +} g_2 	 						\tilde{\Gamma}^* & 
	 -n_{\rm -} \tilde{g_1}^*  	             	\tilde{\Gamma}^* &
    -\left(k^2 + n_{\rm -} \tilde{g_1}^* \right) \tilde{\Gamma}^* & 
     -J B_{\rm J} \lambda \zeta  				\tilde{\Gamma}^* \\[0.3cm]
	  i~\tilde{\alpha} n_{\rm +} \delta \lambda \tilde{B_{\rm J}} &
	  i~\tilde{\alpha} n_{\rm +} \delta \lambda \tilde{B_{\rm J}} & 
	 -i~\tilde{\alpha} n_{\rm -} \delta \lambda \tilde{B_{\rm J}} &
	 -i~\tilde{\alpha} n_{\rm -} \delta \lambda \tilde{B_{\rm J}} & 
     -i~\tilde{\alpha}
\end{pmatrix}
\end{align}
\end{widetext}
%
Where $\tilde{g_1} = g_1 - i \gamma_{\rm NL}$, $\tilde{\Gamma} = (1+ i \Gamma )^{-1}$, $\tilde{\alpha} = \alpha/\tau_{\rm M}$, $\tilde{B_{\rm J}} = B_{\rm J}^\prime / JB_{\rm J}$ and $B_{\rm J} = B_{\rm J}\left(\delta \lambda (n_{\rm +} - n_{\rm -} ) \right)$.
Figures~\ref{fig:T-L} and \ref{fig:gamma-L} show the numerical solution of this eigenvalue problem in parameter space. Stable configurations, for which all $\omega_k$ have a negative imaginary part, are marked with cyan color, while unstable ones with color that represents the fastest rate of the instability (the largest imaginary part of $\omega_k$).

Using the method of analysis of zeros of the corresponding polynomial~\cite{Ostrovskaya_DarkSolitons}, we calculate analytically the  stability condition (see Appendix~\ref{app:Bgv}) 
\begin{gather}
	\lambda^2 B_{\rm  J}^\prime \left( \frac{g_{\rm M} \mu_{\rm B}}{2 k_{\rm B} T} J \lambda (n_{\rm +} - n_{\rm -})\right) < \frac{(g_{1} - g_{2}) k_{\rm B} T}{n_{\rm M} g_{\rm M}^2 \mu_{\rm B}^2 J^2} \label{eq:condition2s}
\end{gather} 
%
and compare it with the analogous condition in the fully polarized case~\cite{Mietki_MagneticPolaron}
\begin{gather}
	\lambda^2 B_{\rm J}^\prime \left( \frac{g_{\rm M} \mu_{\rm B}}{2 k_{\rm B} T} J \lambda n_{\rm +}\right) < \frac{2 g_{1} k_{\rm B} T}{n_{\rm M} g_{\rm M}^2 \mu_{\rm B}^2 J^2} \label{eq:condition1s}
\end{gather} 
Notice the factor of two in the nominator on the right hand side of the above equation. These conditions do not depend on the energy relaxation $\Gamma$, the fact that is reproduced in Fig.~\ref{fig:gamma-L}, and weakly depend on the polariton density. Note that condition (\ref{eq:condition2s}) is valid in the case when $(g1 + g2)>0$, which is always satisfied in polariton condensates. 

The discontinuity of the stability threshold in Figs.~\ref{fig:T-L} and~\ref{fig:gamma-L} is caused by the transition from the polarized to non-polarized regime and the reduction of the number of degrees of freedom for the excitations. Indeed, in the fully polarized case, the stability threshold is given by Eq.~(\ref{eq:condition1s}) while in the partially polarized case a stronger condition Eq.~(\ref{eq:condition2s}) should be taken into account. In result, the system becomes unstable at weaker coupling $\lambda$. 
The ratio of the critical values of the ion-polariton coupling constants in the two cases ($\lambda_{\rm F}$ for fully and $\lambda_{\rm P}$ for partially polarized) can be estimated as
%
\begin{gather}
	\frac{\lambda_{\rm F}}{\lambda_{\rm P}} = \sqrt[]{ \frac{2g_1}{g_1 - g_2} }
\end{gather}
in the limit of small $\lambda$ when the derivative of the Brillouin function is roughly constant. As in the realistic system the intercomponent interaction constant $g_2$ is much smaller than intracomponent interaction constant $g_1$, this leads to a roughly $\sqrt{2}$ jump of the stability threshold. Physically, this reduced threshold for stability is related exactly to the appearance of a new inhomogeneous state of ``polaron lattice'' in the partially spin-polarized regime.
In Appendix~\ref{app:Adr}, we show, in the adiabatic regime, the physical origin of this conditions. 

%
\begin{figure}
	\includegraphics[width=0.5\textwidth]{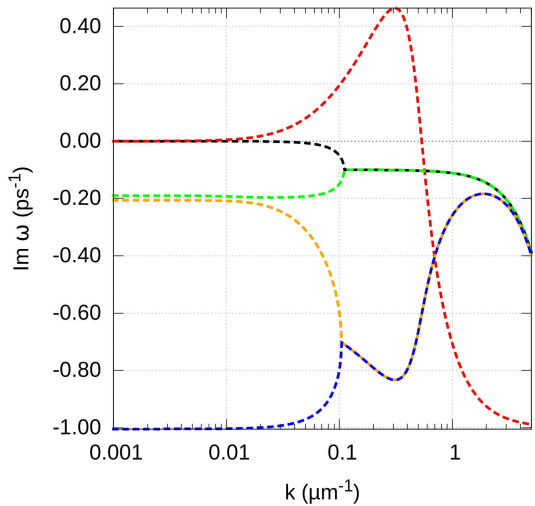}
    \caption{Imaginary part of eigenfrequencies $\omega_k$ of Bogoliubov quasiparticles, for parameters correspoding to Fig.~\ref{fig:1d2s}(a). }
    \label{fig:bogoliubov}
\end{figure}
%
A plot of imaginary parts of eigenfrequencies of the Bogoliubov quasiparticles is shown in Fig.~\ref{fig:bogoliubov}. It corresponds to the simulated evolution presented in Fig.~\ref{fig:1d2s}(a). 
In the contrast to the full polarized case~\cite{Mietki_MagneticPolaron}, the spectrum has five branches instead of three.
Two additional branches (green and black lines in Fig.~\ref{fig:bogoliubov}) appear in the partially polarized case, therefore they correspond to counter-polarized polaritons.
Red branch has values above zero, which evidences dynamical instability of the condensate.

%
%
%
%
\section{\label{sec:conclusions}Conclusions}
%
In conclusion, we investigated a partially polarized exciton-polariton condensate in a semimagnetic semiconductor microcavity.
In a system which is far from equilibrium, we demonstrated several regimes of dynamics. We observed numerically stable solutions, polaron lattice fromation with antiferromagnetic arrangement, and spin-polarized polaron regime.
The lattice regime is paritcularly significant for experiments, since it can be distinguished in a straightforward way from density fluctuations trapped on defects of the semiconductor microcavity. 
We derived a critical condition for the formation of polarons which is different from the one predicted in the fully polarized case.

%
\acknowledgments
%
We thank Alexey Kavokin, Barbara Pi\k{e}tka and Jacek Szczytko for useful discussions. We acknowledge support from the National Science Center grants 2015/17/B/ST3/02273 and 2016/22/E/ST3/00045. 
%
\appendix
%
\section{Homogeneous stationary states in the weak and strong magnetic field limits} \label{app:Wmf}
In this Appendix we calculate the partial-full polarization boundary in the limits of weak and strong magnetic field.
In the weak field limit the Brillouin function can be linearized
\begin{gather}
\begin{aligned}
	M = J &B_{\rm J}\left(\delta \lambda (n_+\!-n_-) + 2 \delta B_0 \right)  \label{eq:Mweak} \\ 
    &\approx c\, \delta \lambda (n_+\!-n_-) + 2 c\, \delta B_0 
\end{aligned}
\end{gather}
where $c = J(J+1)/3$.
From the condition for the two-component stationary state, Eqs.~(\ref{eq:Im1}) and~(\ref{eq:Im2}), we can estimate the value of $n_+$ at the full-partial polarization boundary by substituting $n_- = 0$
\begin{align}
	n_+ = \frac{ 2P_{\rm eff} }{ \Gamma g_1 + \Gamma g_2 + \gamma_{\rm NL} }  \label{eq:n+}
\end{align}
By substituting $M$ and $n_+$ into Eq.~(\ref{eq:Im1}) we get a quadratic equation for $\Gamma$
\begin{align}
	\frac{1}{\Gamma^2} \alpha 
  - \frac{1}{\Gamma}   \left( \beta + \gamma B_0 \right)
  - \varepsilon B_0 = 0
\end{align}
where
$\alpha = P_{\rm eff} \gamma_{\rm NL}$, 
$\beta=P_{\rm eff} \left( 2 \zeta \lambda^2 \delta c - g_1 + g_2  \right)$, 
$\gamma = 2 \zeta \lambda \delta c  \gamma_{\rm NL}$, 
$\varepsilon = 2 \zeta \lambda \delta c \left( g_1 + g_2 \right)$.
Note that $\alpha$, $\gamma$ and $\varepsilon$ are positive.
The appropriate solution is given by
\begin{align}
	\frac{1}{\Gamma} = \frac{\beta + \gamma B_0}{2\alpha} + \frac{\sqrt[]{(\beta + \gamma B_0)^2 + 4\alpha\varepsilon B_0}}{2\alpha} 
\end{align}
When the magnetic field is strong, the magnetization is saturated and the Brillouin function attains the maximum value of unity
\begin{align}
	M = J~B_{\rm J}\left( \delta \lambda (n_+ -n_-) + 2 \delta B_0 \right) = J.
\end{align}
We can obtain the equation for $1/\Gamma$ by putting $M$, $n_+$ into Eq.~(\ref{eq:Im1})
\begin{gather}
\begin{aligned}
	\frac{1}{\Gamma^2} P_{\rm eff} \gamma_{\rm NL} 
  - \frac{1}{\Gamma} \left( \zeta \lambda J \gamma_{\rm NL} - P_{\rm eff} (g_1 -g_2) \right)& \\
  - \zeta \lambda J (g_1 + g_2)& = 0
\end{aligned}
\end{gather}
The positive solution for $1/\Gamma$ does not depend on $B_0$ 
\begin{gather}
	\frac{1}{\Gamma} = \frac{\zeta \lambda J \gamma_{\rm NL} - P_{\rm eff} (g_1 -g_2)}{2 P_{\rm eff} \gamma_{\rm NL} } \\
    + \frac{\sqrt[]{\left(\zeta \lambda J \gamma_{\rm NL} - P_{\rm eff} (g_1 -g_2)\right)^2 + 4 P_{\rm eff} \gamma_{\rm NL} \zeta \lambda J (g_1 + g_2) }}{2 P_{\rm eff} \gamma_{\rm NL}} \nonumber
\end{gather}
and for $g_2 \ll g_1$ can be estimated as
\begin{align}
	\frac{1}{\Gamma} = \frac{\zeta \lambda J  }{P_{\rm eff}}.
\end{align}

\section{Bogoliubov analysis}  \label{app:Bgv}
Determining the condition (\ref{eq:condition2s}) consists of solving the eigenvalue problem $Q_k U_k = \omega_k U_k$ with Bogoliubov matrix (\ref{eq:Qmatrix})
\begin{align}
	\det L_k = \det ( Q_k - \mathds{1} \omega ) = 0
\end{align}

Analyzing the solutions in the limits $k \to 0$ and $k \to \infty$ reveals two (in the case of partial polarization) or three (full polarization) solutions of $\Im (\omega) = 0$ at $k = 0$ and five negative solutions in $k \to \infty$ limit. 
It turns out that analogously to~\cite{Mietki_MagneticPolaron,Ostrovskaya_DarkSolitons} only the purely imaginary branch may have positive imaginary part of the frequency (the red branch in Fig.~\ref{fig:bogoliubov}).
Similar as in~\cite{Mietki_MagneticPolaron,Ostrovskaya_DarkSolitons}, we find the zero-frequency crossing of $\Im (\omega)$ as a function of $k$. 
Since  $\Re (\det L_k) = 0$ we consider the $\omega_1 = 0$ solution and substitute it into $\Im (\det L_k)$ to obtain
\begin{gather}
\begin{aligned}
k^8 + 2 k^6 (n_+ + n_-) ( g_1 - B_{\rm J}^\prime  \delta \lambda^2 \zeta )& \\
+ 4 k^4 n_+ n_-\!\left[(g_1^2 - g_2^2) - 2 B_{\rm J}^\prime \delta \lambda^2 \zeta (g_1 + g_2) \right]& = 0
\end{aligned}
\end{gather}
Apart from $k = 0$ solutions we get
\begin{widetext}
\begin{align}
k^2 = (n_+ + n_-) (B_{\rm J}^\prime \zeta \delta \lambda^2 - g_1 ) 
\pm \sqrt{(n_+ + n_-)^2 (B_{\rm J}^\prime \zeta \delta \lambda^2 - g_1)^2 - 4 n_+ n_- \left( (g_1^2 - g_2^2) - 2 B_{\rm J}^\prime \delta \lambda^2 \zeta (g_1 + g_2) \right)} \label{eq:k2cond}
\end{align}
\end{widetext}
Condensate is stable only if there is no zero crossing of $\Im (\omega)$ as a function of $k$, for $k^2>0$. This is the case when the right hand side of Eq.~(\ref{eq:k2cond}) is less then zero. Otherwise, a range of $k$ with positive imaginary part must exist. 
It is easy to check that the expression under the square root on the right-hand side of (\ref{eq:k2cond}) is always positive.
Considering the solution with the plus sign leads to the condition
\begin{gather}
B_{\rm J}^\prime \lambda^2 < \frac{g_1 - g_2}{2 \delta \zeta} 
\end{gather}
This condition is more restrictive than 
$B_{\rm J}^\prime \zeta \delta \lambda^2 < g_1$ derived for the fully polarized case, 
which is due to the presence of $n_-$ component. The above formula is rewritten in physical units in~(\ref{eq:condition2s}).

\section{Adiabatic approximation} \label{app:Adr}
In the adiabatic approximation we assume that the spin relaxation time $\tau_{\rm M}$ is much shorter than other timescales in the system, and consequently $M=\langle M \rangle$.
By expanding the Brillouin function up to the first order around the stationary value $B_{\rm J}(\delta\lambda \,\Delta n)$ where  $\Delta n = n_{\rm +}\!-\!n_{\rm -}$, we get
\begin{gather}
\begin{aligned}
&M(x,t) = J B_J (\delta \lambda (|\psi_+|^2\!-\!|\psi_-|^2)) \\ 
\approx M_0 &+ J \delta \lambda (|\psi_+|^2\!-\!|\psi_-|^2 ) B_J^\prime \label{eq:Mag}
\end{aligned}
\end{gather}
where we used the notation $M_0\!=\!J B_{\rm J}\!-\!J\delta\lambda \,\Delta n B_{\rm J}^\prime$, $B_{\rm J}^\prime = B_{\rm J}^\prime(\delta\lambda \,\Delta n)$.
Substituting Eq.~(\ref{eq:Mag}) to the dimensionless form of the complex Ginzburg-Landau equation Eq.~(\ref{eq:CGLE2}) leads to
\begin{gather}
i (1+ i\Gamma) \frac{\partial \psi_{\sigma}}{\partial t} =  -\frac{\partial^2  \psi_\sigma}{\partial x^2} \nonumber \\
+ \left( g_1 - \beta B_{\rm J}^\prime \right) |\psi_\sigma|^2 \psi_\sigma  
+ \left( g_2 + \beta B_{\rm J}^\prime \right) |\psi_{-\sigma}|^2 \psi_\sigma  \\
+ i \left( P_{\rm eff} - \gamma_{\rm NL} |\psi_\sigma|^2 \right) \psi_\sigma - \sigma \zeta\lambda M_0 \psi_\sigma \nonumber
\end{gather}
where $\beta = J\zeta\lambda^2\delta$.

We now investigate the stability of the stationary state in the limit of low kinetic energies by a method alternative to the Bogoliubov approximation. The effective potential for the  $\sigma$ component is
\begin{gather}
U_\sigma = ( g_1 - \beta B_{\rm J}^\prime) n_\sigma + (g_2 + \beta B_{\rm J}^\prime) n_{-\sigma} - \sigma\zeta\lambda U_0 \label{eq:potenial_D}
\end{gather}
We consider slight local changes of densities $\Delta n_+$ and $\Delta n_-$ assuming that the value of the derivative of the Brillouin function remains approximately the same. Our question is whether such local fluctuations will have the tendency to grow in time or if they will decay. We consider slow, almost stationary dynamics so assume that the chemical potentials remain practically unchaged
\begin{gather}
0 \approx ( g_1 - \beta B_{\rm J}^\prime) \Delta n_+ + (g_2 + \beta B_{\rm J}^\prime) \Delta n_- \label{eq:chpotcon1_D} \\
0 \approx ( g_1 - \beta B_{\rm J}^\prime) \Delta n_- + (g_2 + \beta B_{\rm J}^\prime) \Delta n_+ \label{eq:chpotcon2_D}
\end{gather}
We inspect how the change of $\Delta n_+$ affects the potential $U_+$. 
The positive value of $\Delta U_+ / \Delta n_+$ corresponds to a stable condensate since the polariton effective mass is positive. Negative value of $\Delta U_+ / \Delta n_+$ means that the denisty fluctuation creates an effectively attractive potential which leads to further density growth, leading to instability.
Combining Eqs.~(\ref{eq:potenial_D}) and~(\ref{eq:chpotcon1_D}) we obtain
\begin{gather}
\frac{\Delta U}{\Delta n_+} = ( g_1 - \beta B_{\rm J}^\prime) - ( g_1 - \beta B_{\rm J}^\prime) > 0 \label{eq:U/n}
\end{gather}
which leads to the stability condition in the fully polarized case $\beta B_{\rm J}^\prime < g_1$ that is equal, in physical units, to Eq.~(\ref{eq:condition1s}).
On the other hand, from Eqs.~(\ref{eq:potenial_D}) and~(\ref{eq:chpotcon2_D}) we get
\begin{gather}
\frac{\Delta U}{\Delta n_+} = ( g_1 - \beta B_{\rm J}^\prime) - \frac{ (g_2 + \beta B_{\rm J}^\prime)^2 } {g_1 - \beta B_{\rm J}^\prime} > 0 \label{eq:U/n2}
\end{gather}
which leads to the condition
\begin{gather}
( g_1 - g_2 - 2 \beta B_{\rm J}^\prime) (g_1 + g_2) > 0 
\end{gather}
Since in a polarton gas we have $g_1+g_2>0$, the condition for stability is
\begin{align}
\beta B_{\rm J}^\prime < \frac{g_1 - g_2}{2} \label{eq:condD},
\end{align}
which corresponds to (\ref{eq:condition2s}) in physical units.

%
\bibliography{bibliography}

\begin{thebibliography}{59}
\bibitem [{Par()}]{Parameters}%
  \BibitemOpen
  \href@noop {} {}\bibinfo {note} {$n$ = $\mathrm{3.91 \times 10^7}$
  $\mathrm{m^{-1}}$, $\tau$ = $\mathrm{ 10^{-12}}$ $\mathrm{s}$, $g_{\rm 1}$ =
  $\mathrm{ 1.28 \times 10^{-9} }$ $\mathrm{meV~m}$, $g_{\rm 2}$ = $\mathrm{
  -6.4 \times 10^{-11}}$ $\mathrm{meV~m}$, $n_{\rm M}$ = $\mathrm{ 1..2 \times
  10^{13} }$ $\mathrm{m^{-1}}$, $B$ = $\mathrm{ 0.001~T }$, $T$ = $\mathrm{
  0.1~K}$, $\lambda$ = $\mathrm{ 2.3 \times 10^{-11}}$ $\mathrm{T~m}$,
  $\gamma_{\rm NL}$ = $\mathrm{ 3.37 \times 10^{-9}}$ $\mathrm{meV~m}$, $P_{\rm
  eff}= P - \frac{1}{2} \gamma_{\rm L} $ = $\mathrm{ 0.0658~meV}$, There is set
  5 meV Rabi splitting and zero photon-exciton detuning; photon effective mass
  $m_{\rm C}$ = $\mathrm{10^{-5}}~ m_{\rm E}$ and exciton effective mass
  $m_{\rm X}$ = $0.22~ m_{\rm E}$.}\BibitemShut {Stop}%
\end{thebibliography}
%
\end{document}